\documentclass[conference]{IEEEtran}
\IEEEoverridecommandlockouts
\usepackage{cite}
\usepackage{amsmath,amssymb,amsfonts}
\usepackage{algorithmic}
\usepackage{graphicx}
\usepackage{textcomp}
\usepackage{xcolor}
\usepackage{comment}
\usepackage[utf8]{inputenc} 
\usepackage{tikz,pgfplots}
    \usepgfplotslibrary{dateplot}
    \usepgfplotslibrary{fillbetween}
    \usepgfplotslibrary{colorbrewer}
    \usetikzlibrary{shapes.multipart}
    \usetikzlibrary{automata,positioning}

\def\BibTeX{{\rm B\kern-.05em{\sc i\kern-.025em b}\kern-.08em
T\kern-.1667em\lower.7ex\hbox{E}\kern-.125emX}}

\newcommand{\subparagraph}{}
\usepackage{titlesec}

\begin{document}

\title{Energy Communities: From European Law to Numerical Modeling
}

\author{\IEEEauthorblockN{ Clément Alaton\textsuperscript{1}, Jesus Contreras-Ocaña\textsuperscript{2}, Philippine de Radiguès\textsuperscript{1}, Thomas Döring\textsuperscript{1}, Frédéric Tounquet\textsuperscript{1}}
\IEEEauthorblockA{\textsuperscript{1}\textit{ENGIE Impact}, \textsuperscript{2}\textit{ENGIE Global Energy Management}\\
\textit{Blvd. Simon Bolivar 34,  Brussels, Belgium} }
}

\maketitle

\begin{abstract}
In 2019, the European Union introduced two new actors in the European energy system: Renewable and Citizen Energy Communities (RECs and CECs). Modelling these two new actors and their effects on the energy system is crucial when implementing the European Legislation, incorporating energy communities (ECs) into the electric grid, planning ECs, and conducting academic research.  This paper aims to bridge the gap between the letter of the law and numerical models of ECs. After introducing RECs and CECs, we list elements of the law to be considered by regulators, distribution system operators, EC planners, researchers, and other stakeholders when modelling ECs. Finally, we provide three case studies of EC models that explicitly include elements of the European Law.
\end{abstract}


\section{Introduction}

In 2019, the European Union (EU) presented in the Clean Energy for all Europeans Package (CEP), a broad set of measures to promote building energy efficiency, renewable energy, consumer rights on self-generation, cross-border cooperation, among others~\cite{CEP}. To accomplish the aforementioned goals, the CEP introduced two new actors in the European energy system: Renewable Energy Communities (RECs) and Citizen Energy Communities (CECs). Their intent is to empower citizens and achieve the following~\cite{jasiakenergy}:
\begin{enumerate}
    \item Increase the public acceptance of new energy projects.
    \item Mobilize private capital for the energy transition.
    \item Increase flexibility in the market.
\end{enumerate}

Both types of energy communities (ECs) have specific membership criteria, governance requirements, and purposes~\cite{roberts2019energy}. In a nutshell, and to be elaborated in Section~\ref{sec:ECs_in_the_CEP}, RECs and CECs are entities
\begin{itemize}
    \item whose membership is open and voluntary and must be controlled by natural persons, small and medium enterprises, and/or local authorities;
    \item whose primary purpose is to provide environmental, economic, or social community benefits rather than financial profits;
    \item who are entitled to engage in activities such as energy production and consumption, sharing electricity, and market participation.
\end{itemize}

As of May 2020, only a few European Member States have implemented the CEP. Some countries have published proposals and consultation documents that preview their implementation approach. Furthermore, frameworks for RECs and CECs will be partly codified in secondary laws and regulations, which are at early stages even in countries with early implementations.

\subsection{Purpose and contributions of this paper}

Relevant power system actors (e.g. regulators, distribution system operators, suppliers, EC planners, etc.) will need to model ECs in their decision-making processes. For example, regulators will need to incorporate intra-community sharing when designing network tariffs; distribution system operators (DSOs) will need to incorporate ECs in their grid expansion plans; suppliers will need to account for local production and intra-community sharing; EC planners will need to incorporate the non-profit nature of RECs and CECs in their operational and financial plans. 

The purpose of this paper is \textit{to provide both the academic community and industrial practitioners with an understanding of European Legislation on ECs and to identify legal provisions relevant to EC modeling}. The contributions of this paper are three-fold.
\begin{itemize}
\item First, a brief overview of RECs and CECs and implementation examples in three EU Member States (Sections~\ref{sec:ECs_in_the_CEP} and~\ref{subsec:Implementations_in_MSs}). 
\item Second, the identification CEP elements relevant to EC modeling (Section~\ref{sec:Modeling_EC_Legislation}).  
\item Three case studies of EC models that integrate elements of the CEP. (Section~\ref{sec:CaseStudies})
\end{itemize}

\subsection{Literature review}

In Section~\ref{sec:ECs_in_the_CEP}, we provide a brief overview of RECs and CECs. However, the interested reader is invited to consult References~\cite{roberts2019energy, frieden2019overview, gancheva2018models, caramizaru2020energy} for deeper analysis of the definitions and emerging frameworks of ECs.      In~\cite{roberts2019energy}, Roberts et al. analyse the definitions, similarities, and differences of both RECs and CECs, and address some common questions. The work in~\cite{frieden2019overview} analyses emerging EC and collective-self consumption regulatory frameworks in Europe including those of Austria, Spain, Belgium, and Germany. In~\cite{gancheva2018models}, Gancheva et al. present an overview of emerging local energy ownership models in the EU and provide legal and regulatory framework recommendations. Similarly, Caramizaru and Uihlein identify different types of ECs (in terms of drivers, organizational forms, activities, etc.) and provide policy recommendations~\cite{caramizaru2020energy}. 

Several works have studied the complex relationship between ECs, the electric system, and the regulatory framework in which they develop. Both works in~\cite{brummer2018community} and~\cite{koirala2016energetic} study the barriers, benefits, and trends and how policy schemes affect the development of ECs. In~\cite{abada2018unintended} Abada et al. show how inadequate EC network tariffs can lead to an excess EC adoption and non-desirable outcomes for the grid.

The work in~\cite{mendes2011planning} provides an overview of tools for planning and analysis of ECs (e.g., HOMER and DER-CAM). Torabi Moghadam et at. provide a general procedure and process to create an EC that is divided into the following three steps: building identification, feasibility analysis, and target group involvement~\cite{torabi2020mainstreaming}. The work in~\cite{schwarz2019agent} present an agent-based model for EC and DSO dispatch coordination. In~\cite{schram2020trade}, the authors present an EC operation model that incorporates both economic and environmental objectives. Regarding ``real world'' examples of ECs, reference~\cite{reijnders2020energy} presents a control strategy for the Dutch GridFlex Heeten EC and~\cite{jereb2019energy} introduces the Energy Lu\v{c}e in Slovenia.

\section{Energy Communities in the Clean Energy Package}
\label{sec:ECs_in_the_CEP}

\subsection{Renewable and Citizen's Energy Communities}
\label{subsec:RECs_and_CECs}

The CEP defines two types of ECs. The recast of the Renewable Energy Directive (REDII) introduces RECs and the Internal Electricity Market Directive (IEMD) introduce CECs~\cite{red2018directive, IEMD2019directive}. While RECs and CECs are broadly similar entities they have important differences regarding
\begin{itemize}
    \item their entitled activities and
    \item their participation criteria and governance models.
\end{itemize}
Both forms of ECs, however, must aim at providing environmental, economic, and social benefits to its members and/or the places where they operate rather than mainly pursuing profit. 

\subsubsection{Activities of ECs}

As their name suggests, RECs are restricted to renewable energy-related activities. In addition to electricity, RECs can engage with diverse energy carriers such as heat or bio-gas. CECs, on the other hand, can be involved with non-renewable energy but are limited to the electricity sector. 

The set of activities CECs are entitled to engage in is broader than those of RECs. While the REDII entitles RECs to produce, consume, sell, and access all suitable markets, the IEMD entitles CECs to participate in distribution, aggregation, energy efficiency, and energy services, in addition to activities of RECs. Additionally, members of both types of ECs are entitled to share locally produced electricity.

\subsubsection{EC participation and governance}
\label{subsubsec: EC_part_gov}

The Directives require participation in REC and CEC to be open and voluntary. In other words, applicants cannot be arbitrarily refused membership and members should have the right to a fair exit procedure. 

REC membership is restricted to natural persons, small and medium enterprises (SMEs) and local authorities. Furthermore, the REC should be controlled by members located in the proximity of the community's renewable energy projects. On the other hand, membership to CECs is not restricted nor are the members who control the CEC geographically constrained. However, the IEMD denies medium and large enterprises the right to control a CEC. For both REC and CEC, companies whose main activity is in the energy sector are not allowed to exert control over the community. 

A peculiarity of RECs is the autonomy provision. That is, the REDII requires RECs to be independent of the private interests of any member or non-member. In other words, no member should be able to favour its interest over those of the community nor any external entity should exert control of the REC. 

\subsubsection{Enabling framework for ECs}
 The Directives mandate the European Member States to establish a regulatory framework and remove administrative and regulatory barriers to the development of ECs. The framework must, among others, 
\begin{itemize}
    \item facilitate access to finance and information,
    \item  secure the rights and  non-discriminatory treatment of members,
    \item ensure proportionate and transparent procedures (e.g. registration and licensing),
\end{itemize} 
 
\subsection{Renewable self-consumers and active customers}
\label{subsec:CSC}

The Directives define renewable self-consumers and active customers, two concepts related to ECs. The REDII defines renewable self-consumers as customers who generate, self-consume, store, and sell renewable electricity locally (e.g. at their home or apartment building). An analogue concept in the IEMD is the active customer. These are final customers who consume, store or sell electricity produced locally. Additionally, active customers are entitled to provide flexibility and energy efficiency services.  For both renewable self-consumers and active customers, said activities should not be their primary commercial or professional endeavours.

Like ECs, the intention of renewable self-consumers and active customers is to empower citizens and make them active market actors. However, RECs and CECs are more complex legal and social concepts that are allowed a greater geographical extension, a broader set of activities, and are required to have special governance structures. Self and active consumption, on the other hand, are contained to smaller premises and do not require the formation of a legal entity (some countries have imposed this restriction, e.g. France).

\section{Implementations in European Member States}
\label{subsec:Implementations_in_MSs}

In this section, we introduce examples of implementations of EC legislation in the European Member States. We discuss the implementation of the proximity requirement for RECs in Wallonia (Belgium) and the collective self-consumption framework and tariffs implemented in France. 

\subsection{The proximity requirement of RECs in Wallonia}
\label{subsec:prox_Wallonia}
Wallonia is the southern french-speaking region of Belgium. While the Belgian government has some jurisdiction on matters of energy, the Regional governments are leading the task of implementing the CEP.

As provided by the CEP, RECs are required to be controlled by members ``located in the proximity of the renewable energy projects that are owned and developed by [the REC].'' Three general approaches to defining ``proximity'' are observed in Europe: distance, network (e.g. voltage level) and administrative (e.g. administrative regions)-based approaches.

An interesting implementation of the proximity criterion is provided by the Walloon Decree of May 2\textsuperscript{nd}, 2019 which defines the  ``local perimeter'' as the perimeter where electricity off-take and injection points are below one or more medium/low voltage transformers (i.e. a network-based definition). Additionally, the Decree gives the regulator power to further restrict the perimeter in a way “that mobilizes the technically, socially, environmentally, and economically optimal portion of the network with the view of promoting local collective self-consumption of electricity”. This approach aims to tailor the proximity criterion on a case-by-case basis to advance the environmental and social goals of the Directives. However, a downside of granting the regulator this kind of discretion is that it generates uncertainty that EC developers and DSOs will have to cope with.  

\subsection{Energy allocation for collective self-consumption in France}

The French self-consumption Decree of the 28\textsuperscript{th} of April 2017 defines rules for the activity of energy sharing and self-consumption and establishes a procedure to communicate the repartition of collectively self-consumed energy to the party responsible for metering (the DSO in France). The collective self-consumers must provide the DSO with the allocation of local energy production among the consumers on a 30-minute basis\footnote{Alternatively, the self-consumers can also indicate the computation method or rules.}. Communicating the energy repartition is mainly for billing purposes as it is needed to calculate the residual energy that each of the consumers received from their supplier. 

 This Decree gives the self-consumers the freedom to allocate locally produced energy as they please (as long as the energy allocated to a consumer at each time period does not exceed their load). However, a default rule of repartition is available and automatically applied if communication between the self-consumers and the DSO is interrupted or if the members do not define a custom rule of repartition. The default rule allocates local production in proportion to their load. 
 
 The Decree aims to ensure cooperation between DSO and the collective self-consumers by establishing a simple and seamless communication procedure for energy sharing and self-consumption. Additionally, the free choice of the allocation rule gives the self-consumers the opportunity to tailor their activities to favour environmental, social, and economic benefits (e.g. by allocating energy on solidarity rather than on a commercial basis). Finally, the decree enables less experienced and knowledgeable consumers to engage in collective self-consumption via the default rules.

\subsection{Grid tariffs for energy sharing in France}

France has been among the first Member States to adopt dedicated grid tariffs for collective self-consumption.  In deliberation of June 25\textsuperscript{th} 2019, the French energy regulator (CRE) approved the set of tariffs and fees that apply to collective self-consumers~\cite{TURPE5bis}.

The tariff structure discriminates based on the source of the electricity. Network fees for locally produced consumption are significantly lower than those for ``regular'' electricity from the grid. The rationale for this difference is that locally produced energy use a smaller portion of the network while regular grid electricity typically travels long distances across the network. Naturally, this distinction is meant to incentivize consumption of locally-produced energy.

\section{Modeling the Energy Community Legislation}
\label{sec:Modeling_EC_Legislation}

This section presents a non-exhaustive list of the provisions and topics in the REDII and the IEMD that will likely be incorporated in EC models by regulators and practitioners. For each of the topics, we briefly introduce the challenges that would need to be addressed partly by modelling the provisions in the Directives.

\subsection{Purpose of the community}
The purpose of RECs and CECs should be to ``provide environmental, economic or social community benefits ... rather than financial profits''~\cite{red2018directive, IEMD2019directive}. This may require regulators to develop models to quantify benefits such as emission reductions (e.g. by assigning a social cost of carbon), reactivation of local economies, or the inclusion of disadvantaged households in the energy transition. EC planners, on the other hand, may need to develop plans and control algorithms that account for those benefits (e.g. via multi-objective optimization). 

\subsection{Market access}

The Directives entitle ECs to ``access all suitable energy markets both directly or through aggregation in a non-discriminatory manner." Regulators may need to address this provision and make market design changes to avoid discrimination of ECs and other new players in the energy system. DSOs may need to update codes and regulations to allow EC market access. The communities themselves may need to develop market bidding models whether or not to access the market via an aggregator.

\subsection{Electricity sharing}
\label{subsec:Elect_Sharing}
An important element that both the REDII and the IEMD share is that they entitle members of RECs and CECs to ``share ... renewable energy that is produced by the production units owned by that [EC].'' Regulators and DSOs face the challenge of establishing frameworks that enable sharing and designing financially sustainable network tariffs that, among others, recuperate network costs and avoid unintended cross-subsidies. ECs, on the other hand, face operational and planning challenges (e.g. optimization of resources in the presence of peer-to-peer sharing).

\subsection{Management of distribution networks}
The IEMD gives the European Member States the possibility of allowing CECs to ``manage distribution networks in their area of operation and establish the relevant procedures.'' This provision has the potential to give ECs significantly greater autonomy and self-determination. However, it gives regulators and DSOs with significant regulatory and technical challenges (e.g. how to transition to a more fragmented distribution network). It would also imply technical (e.g. voltage management, maintenance, and operation) and financial challenges (e.g. how to select and fund network investments) for ECs.

\subsection{Geographical restrictions}
As mentioned in Section~\ref{subsubsec: EC_part_gov}, the Directives require members who control RECs to be located in the ``proximity'' of the community's renewable energy projects. Regulators, on one hand, have the task of defining ``proximity.'' As Section~\ref{subsec:prox_Wallonia} discusses, three approaches to define proximity are currently observed in Europe: distance, network, and administrative-based methods.

\subsection{Governance}

ECs face a challenge additional to those faced by ``traditional'' microgrids: their governance system should be, in a sense, democratic (as per the autonomy requirement mentioned in Section~\ref{subsubsec: EC_part_gov}). Regulators and researchers may consider using tools like game theory to understand the decision-making process of ECs and foresee their impact on the system. An example of the use of such tools for ECs is presented in~\cite{abada2020viability} where the authors use cooperative game theory to test the ability of ECs to adequately share gains.

\subsection{Regulatory framework and network tariffs}

The Directives may require the European Member States to reassess relevant regulatory and administrative frameworks (e.g. licensing, interconnection, and market participation rules) to ensure unjustified barriers are removed. Additionally, regulators and/or DSOs will likely need to revise network tariffs and adapt them for ECs.  Since consumption and production patterns of ECs, and consequently the marginal cost they impose on the system, will likely differ from those of ``regular'' consumers, the appropriate tariff should also differ.

\section{Case Studies}
\label{sec:CaseStudies}

\subsection{EC design in Wallonia, Belgium}

Here we present a sizing and operation model of a prospective 13-participant EC in Wallonia. The model is a mixed-integer linear program (MILP) that minimizes investment (of a solar plant and an energy storage system)  and operation costs and incorporates elements of CEP that include:
\begin{itemize}
   \item The pursuit of social, economic, and environmental benefits by ECs. 
    \item The definition of electricity sharing, self-consumption and their associated tariffs. 
\end{itemize}
The model was implemented in PROSUMER, a microgrid and EC planning tool developed in-house at ENGIE Impact. We also include the CEP-granted right that community members should be given equal treatment (compared to non-members) by suppliers but do not discuss it due to paper limits. 

To assess the social, economic, and environmental benefits of the EC, we compare the EC with the business-as-usual situation using the following indicators: total energy cost, the share of renewable energy use, and the impact on relevant stakeholders. Specifically, we study the EC's impact on the DSO's revenue from network tariffs and the traditional supplier's revenue from energy sales.

\begin{figure}
\centering

\includegraphics[width=0.5\textwidth]{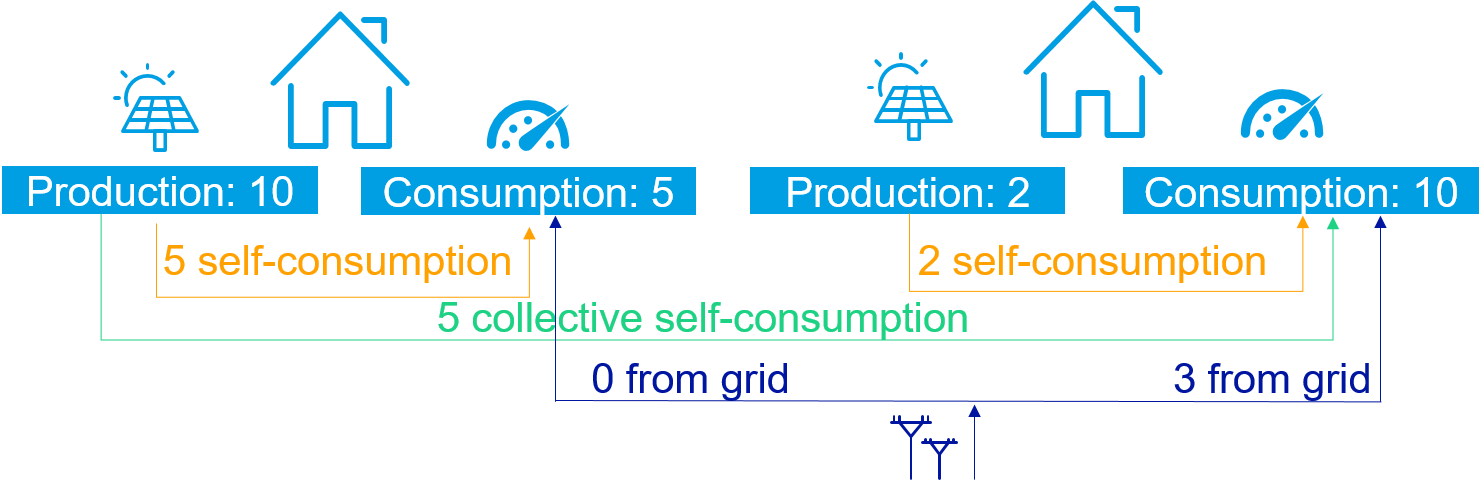}

  \caption{The PROSUMER EC model distinguishes from individual self-consumption, collective self-consumption, and consumption from the grid. These distinctions are important to calculate the financial impact that different tariff and tax designs can have on the different stakeholders related to an EC (participants, the community as a whole, the DSO and the traditional energy supplier).}
  \label{fig:EC_mechelen}
\end{figure}

As discussed in Section~\ref{subsec:Elect_Sharing}, the Directives entitle ECs to ``share ... renewable energy that is produced by the production units owned by that [EC].'' PROSUMER implements the requirement that shared energy must be produced in the EC using linear constraints and optimization variables. Likewise, the model distinguishes between self-consumption, shared energy, and energy from the grid. This distinction is important as each source of energy is associated with different costs, tariffs, and taxes. Figure~\ref{fig:EC_mechelen} illustrates the different energy sources in the PROSUMER model.

\subsection{Collective self-consumption in France}

This case study is based on the work from Contreras-Oca\~na et al. in ~\cite{contreras2020integrated}. In it, the authors propose an integrated framework for long-term planning (i.e. sizing of assets) and short-term operation of an EC with shared solar and energy storage facilities.

The framework proposed in~\cite{contreras2020integrated} incorporates the ``key of repartition'' requirement of the French collective self-consumption framework. A key of repartition is the set of information that indicates to which member(s) local EC generation is assigned to. 
The goal of this case study is to showcase how the key of repartition requirement can be incorporated into long and short-term planning models of ECs.

In~\cite{contreras2020integrated}, a key of repartition is modelled as a matrix $G$ of size $T \times N$, where $T$ is the number periods of energy allocation and $N$ is the number of community members. A key of repartition for a 15-member community who receive energy on a 30-minute basis during 24 hours would be represented by a $48 \times 15$ matrix.

A key of repartition should comply with the physical realities of the system in question. In other words, the values that the matrix $G$ is allowed to take are physically constrained. Namely, the total energy allocated among community members at any period must not be greater than the total generation and the energy allocation to a member must not exceed its load. The model in~\cite{contreras2020integrated} also requires the key of repartition to allocate all generation among community members if the aggregate load is greater than the generation (i.e. there is a local generation deficit) and that the aggregate load must be entirely supplied with local energy when there is a local generation surplus. 

Let the vector $g$ of length $T$ denote local generation per time period and the $T\times N$ matrix $L$ the community member load. The physical constraints the key of repartition can be described mathematically by the set
\begin{equation}
    \mathcal G( g,L) = \left \{ G \mid  \; 0 \le G \le L , G1 = \min\left( g,L1 \right) \right\}. \label{eq:key}
\end{equation} 
Here, the inequality $0 \le G \le L $ restricts the energy allocated to each member to be greater than zero but smaller than their load. The equality $G1 = \min\left( g,L1 \right)$ requires the sum of allocated energy at every time period ($G1$) to equal the local generation when local generation is smaller than the aggregate load ($L1$) and to equal the aggregate load when local generation exceeds the aggregate load. 

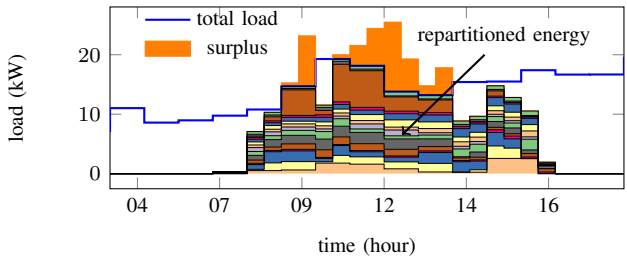
\begin{figure}
\centering
\begin{tikzpicture}

\begin{axis}[
       cycle list/Accent,
  date coordinates in=x,
  ticklabel style = {font=\footnotesize},
  xticklabel=\hour,
  xmin  = 00-01-00 04:00, xmax = 00-01-00 19:00,
  width=.95\linewidth, height=4cm, 
  y label style={ font=\footnotesize},
    x label style={font=\footnotesize},
	ylabel=load (kW), xlabel= time (hour),
		legend style={at={(0.05,0.85)}, draw=none,fill=none,
		anchor=west,legend columns=1, font=\footnotesize},
]
\path[name path=axis] (axis cs:00-01-00 00:00,0) -- (axis cs:00-01-00 23:30,0);
\addplot[const plot,name path=LOAD, color=blue, thick] table [x=time,y=load, col sep=comma] {actual_key_sample.dat};
\addplot[const plot, name path=SOLAR, color=orange, line width=0pt,forget plot] table [x=time,y=production, col sep=comma] {actual_key_sample.dat};
\addplot[const plot,name path=B, color=black, line width=0pt,forget plot] table [x=time,y =x1 , col sep=comma] {actual_key_sample.dat};
\addplot[const plot,name path=C, color=black, line width=0pt,forget plot] table [x=time,y =x2 , col sep=comma] {actual_key_sample.dat};
\addplot[const plot,name path=D, color=black, line width=0pt,forget plot] table [x=time,y =x3 , col sep=comma] {actual_key_sample.dat};
\addplot[const plot,name path=E, color=black, line width=0pt,forget plot] table [x=time,y =x4 , col sep=comma] {actual_key_sample.dat};
\addplot[const plot,name path=F, color=black, line width=0pt,forget plot] table [x=time,y =x5 , col sep=comma] {actual_key_sample.dat};
\addplot[const plot,name path=G, color=black, line width=0pt,forget plot] table [x=time,y =x6 , col sep=comma] {actual_key_sample.dat};
\addplot[const plot,name path=H, color=black, line width=0pt,forget plot] table [x=time,y =x7 , col sep=comma] {actual_key_sample.dat};
\addplot[const plot,name path=I, color=black, line width=0pt,forget plot] table [x=time,y =x8 , col sep=comma] {actual_key_sample.dat};
\addplot[const plot,name path=J, color=black, line width=0pt,forget plot] table [x=time,y =x9 , col sep=comma] {actual_key_sample.dat};
\addplot[const plot,name path=K, color=black, line width=0pt,forget plot] table [x=time,y =x10 , col sep=comma] {actual_key_sample.dat};
\addplot[const plot,name path=L, color=black, line width=0pt,forget plot] table [x=time,y =x11 , col sep=comma] {actual_key_sample.dat};
\addplot[const plot,name path=M, color=black, line width=0pt,forget plot] table [x=time,y =x12 , col sep=comma] {actual_key_sample.dat};
\addplot[const plot,name path=N, color=black, line width=0pt,forget plot] table [x=time,y =x13 , col sep=comma] {actual_key_sample.dat};
\addplot[const plot,name path=O, color=black, line width=0pt,forget plot] table [x=time,y =x14 , col sep=comma] {actual_key_sample.dat};
\addplot[const plot,name path=P, color=black, line width=0pt,forget plot] table [x=time,y =x15 , col sep=comma] {actual_key_sample.dat};
\addplot[color=orange] fill between[of=P and SOLAR];
\addplot fill between[of=B and axis,forget plot];
\addplot fill between[of=C and B,forget plot];
\addplot fill between[of=D and C,forget plot];
\addplot fill between[of=E and D,forget plot];
\addplot fill between[of=F and E,forget plot];
\addplot fill between[of=G and F,forget plot];
\addplot fill between[of=H and G,forget plot];
\addplot fill between[of=I and H,forget plot];
\addplot fill between[of=J and I,forget plot];
\addplot fill between[of=K and J,forget plot];
\addplot fill between[of=L and K,forget plot];
\addplot fill between[of=M and L,forget plot];
\addplot fill between[of=N and M,forget plot];
\addplot fill between[of=O and N,forget plot];
\addplot fill between[of=P and O,forget plot];
\legend{total load, surplus}

\node[anchor=south] (source) at (axis cs:00-01-00 15:30, 20){\footnotesize repartitioned energy};
       \node (destination) at (axis cs: 00-01-00 12:15,5){};
       \draw[->, thick](source)--(destination);
\end{axis}

\end{tikzpicture}

  \caption{Total load, the key of repartition, and surplus for a sample day. The repartitioned energy is represented by the colored area under the blue curve with each color representing a different community member.}
  \label{fig:key_example}
\end{figure}

The mathematical representation of the key of repartition formulated in Eq.~\eqref{eq:key} serves important functions in the long and short-term planning of ECs. In the long-term planning phase, one of the goals is to estimate the amount of local generation that each member receives. In the short-term, the goal is to formulate a key of repartition to allocate, in real-time, local generation to each of the members. The constraints~\eqref{eq:key} allow both models to only consider feasible energy allocations. Fig.~\ref{fig:key_example} shows the key of repartition of locally produced energy for a 15 member EC with solar and energy storage located in the south of France.

\subsection{Som Energia - Collective investment in Spain }
Som Energia is a Spanish cooperative focused in production and supply of renewable energy. An interesting program by Som Energia is the ``Generation kWh'' program which allows individuals to provide Som Energia with an interest-free loan of as little as 100 € for construction of renewable energy projects.  In return, the individuals receive (at cost of production) a share of the generation proportional to their investment. The loan is paid back to the individuals throughout the 25-year operational life of the project~\cite{SomCondiciones}. 

Today, the Generation kWh program operates two PV projects and has one more in the planning phase. The PV projects in operation are the Alcolea del Río plant in Sevilla (2,1 MW peak and a 2 million € investment) and the Fontivsolar in Fontiveros (990 kW peak and an 850.000 € investment). The La Serra PV plant in Anglesola (2,8 MW peak and 1,8 million € investment) is currently in the planning phase. 

The Generation kWh program advances at least two of the main goals of ECs in the CEP. First, it is an innovative tool to mobilize private capital for the energy transition. It ``democratizes'' investing in renewable energy by allowing individuals to make relatively small contributions to the development of PV projects. Second, it has the potential to increase the public acceptance of renewable energy projects. We consider that enabling individual citizens to fund concrete projects (rather contributing than through relatively abstract mechanisms like ``green certificates'') could increase public awareness and involvement and boost support for the energy transition.

\section{Conclusion}

Having an informal status for a long time, energy communities (ECs) are now recognised as full actors of the European energy system. In a context of energy sector decentralisation and digitalization, the economic and social dimensions of ECs further disrupt the conception, modelling, planning, and operation of the energy system.

This paper provides a high-level understanding of the legal definitions of Renewable and Citizen Energy Community concepts and identifies their elements that require modelling from distribution system operators, suppliers, EC planners, regulators, academics, and others. Some of the main elements that we identify are the pursuit of social and environmental benefits (i.e. their purpose), energy sharing, geographical restrictions of RECs, market access, among others.

While some aspects of ECs are commonly modelled (e.g. energy management of a microgrid ), modelling the EC-related legal provisions is still at an early stage. While this paper is not meant to be exhaustive, we aim at shedding light on elements of the law that need to be taken into account by modellers.  In our view, the design of cost-reflective and fair network tariffs for ECs is today one of the most challenging modelling exercises.

\bibliographystyle{IEEEtran}
\bibliography{IEEEabrv,IEEEexample}

\end{document}